 \documentclass[final,1p,times,authoryear]{elsarticle}

\usepackage{amssymb}
\usepackage{lipsum}
 \usepackage{amsmath} 
\usepackage{url,hyperref,lineno,microtype,subcaption}
\usepackage[onehalfspacing]{setspace}
 
\pdfoutput=1

\begin{document}

\begin{frontmatter}



\title{A tabu search-based LED selection approach safeguarding visible light communication systems}


\author[first]{Shi, Ge\,$^{1}$,  Cheng, Wei\,$^{2}$, Gao, Xiang\,$^{2}$, Wei, Fupeng\ $^{1}$, Zhang, Heng\,$^{2}$ , and Wang, Qingzheng$^{1}$}
\author[first]{$^{1}$The School of Information Engineering, North China University of Water Resources and Electric Power, Zhengzhou, 450046, P.R.China. \\
$^{2}$The School of Electronics and Information, Northwestern Polytechnical University, Xi'an, 710129, P.R.China.}

\begin{abstract}
In this paper, we investigate the secrecy performance of a single-input single-output visible light communication (VLC) channel in the presence of an eavesdropper. The studied VLC system comprises distributed light-emitting diodes (LEDs) and multiple randomly located users (UEs) within an indoor environment. A sum secrecy rate maximization problem is formulated to enhance confidential transmission by selecting the optimal LED for each UE. To address the non-convex and non-continuous nature of this problem, we propose a tabu search-based algorithm that prevents entrapment in local optima by organizing the trial vectors from previous iterations. Furthermore, we develop three straightforward LED selection strategies that reduce computational complexity by employing fixed criteria to choose one LED for each UE. We also examine the convergence and complexity analysis of the proposed algorithm and strategies. The results demonstrate that the secrecy performance of our proposed algorithm is very close to the global optimal value and surpasses that of the developed strategies.
\end{abstract}




\end{frontmatter}




\section{Introduction}

Visible light communication (VLC) is an emerging wireless technology that uses visible light to transmit data. VLC works by modulating the intensity of optical signals from light sources to encode and transmit information. VLC has gained significant attention recently due to its potential for fast, secure, and energy-efficient data transmission \citep{SylSurvey}. It can operate where radio frequency (RF) signals are restricted, such as in hospitals and aircraft. VLC also offers transmission security since the signals do not penetrate walls and are not susceptible to electromagnetic interference. VLC has various applications, including indoor positioning\cite{Ge2019}, smart lighting, and high-speed internet access\cite{Shimaa2020}. 

VLC systems have a broadcasting characteristic, which allows multiple users (UEs) to access the information transmitted from every light-emitting diode (LED) that is inside their detective areas. A malicious UE can potentially gather confidential data not intended for them. Traditional upper techniques may not effectively safeguard VLC systems since each transmission requires private key exchanges and management. Fortunately, physical layer security (PLS) has appeared as a promising method to enhance wireless security by utilizing wireless channel characteristics. In PLS, the performance is determined by the transmission performance gap between a legitimate channel and an eavesdropper (Eve) channel \citep{Khisti_MISO}. Therefore, to enhance the security of VLC systems, strategies must be designed to improve the signal-to-interference-plus-noise ratio (SINR) of the legitimate UE while degrading that of Eve. This approach has been extensively investigated in RF systems, but PLS in VLC systems requires special investigation due to their differing characteristics.

Beamforming-based techniques have drawn considerable attention in enhancing the wireless security of VLC systems by exploiting spatial degrees of freedom \citep{Mostafa2015, Arfaoui2017}. In particular, for a multiple-output single-input (MISO) VLC wiretap system, a zero-forcing precoder was adopted as a sub-optimal approach in \citep{Mostafa2015} for secure transmission when the channel state information (CSI) of Eve is available. When the location of a passive Eve is known within a bounded area, robust beamforming was designed to enhance the worst-case secrecy rate by nullifying the positions where the Eve may exist \citep{Arfaoui2017}. In \citep{Ge2022}, the author has studied a sum secrecy rate maximization problem of a MISO VLC broadcast wiretap channel based on non-orthogonal multiple access.

Beamforming-based techniques demonstrate superior performance as they enable transmitters to steer the direction of optical signals and adjust their power levels. By dynamically controlling these parameters, beamforming-based methods can focus the transmitted light toward specific UEs or areas of interest. This targeted transmission improves signal strength, enhances coverage, avoids eavesdropping, and reduces interference from other sources, leading to superior performance in terms of signal reception and reliability. However, beamforming-based methods may require additional equipment, such as lenses and signal processing circuits, to adjust the amplitude and direction of the optical signals to create a beam targeting UEs. VLC systems are typically intensity-modulation/direct-detection systems, which use incoherent light and do not have the capability to directly control the phase of the light wave. Beamforming devices that are designed for coherent optical communication systems, such as those using infrared radiation lasers \citep{Ton2019}, cannot be directly transplanted to VLC systems.

Another promising technique for improving the performance of VLC systems is LED selection. While LED selection methods may not achieve the same level of performance as beamforming-based methods, they offer a compromise between performance and complexity \cite{Cho2020}.
The computational complexity comparison between these two methods can be analyzed based on two factors: solution space and problem dimension. Beamforming-based methods refine solutions in continuous space, while LED selection methods search for solutions in discrete space. Due to its finer-grained solution space, the former methods generally require more iterations compared to the latter methods. 
 Besides, the problem dimension of beamforming-based methods is $M$ times larger than that of LED selection methods, where $M$ represents the number of UEs. In beamforming-based methods, the optimization variables for a secrecy-rate maximization problem are $M$ beamforming vectors that allocate the coefficients from LEDs to each UE, with a size of $1\times K$, where $K$ represents the number of LEDs. On the other hand, in LED selection methods, the optimization variable is one vector that includes the indices of LEDs, with a size of $1\times K$.

To enhance transmission quality in LED selection methods, the optimal criterion often involves maximizing the minimum Euclidean distance. An LED selection algorithm based on the support vector machine of a VLC system was proposed in \citep{ZHANG2021} to improve the bit error rate, where the maximum minimum Euclidean distance is
selected as the key performance index of the system to construct the label vector of the training samples. In \citep{Yang2022}, a hybrid dimming scheme based on LED selection was proposed to maximize the throughput of a multiple-user multiple-cell VLC system. An LED selection subproblem was formulated as a mixed-integer problem and solved using a penalty method. In \citep{Yitian2021}, a group-based LED selection-aided scheme was designed to improve the bit error rate by reducing the effect of channel correlation. 

For security purposes, the main idea behind LED selection is to choose the best combinations between LEDs and UEs that can maximize the data rates of UEs while minimizing that of Eve. 
An LED selection scheme that chooses the nearest LED for UEs was proposed in \citep{Cho2018} to reduce the secrecy outage probability of a VLC system, providing a practical and near-optimal solution. In addition, a ternary scheme investigated in \citep{Cho2019_AN} allowed each LED to select its mode among jamming, transmitting, and silent, reducing the complexity of a joint beamforming and jamming scheme without much loss of secrecy outage probability. 

 LED selection can involve large search spaces due to the number of possible combinations between LEDs and UEs. Decision tree algorithms explore multiple branches and evaluate potential splits at each level, which can lead to increased computational complexity and longer processing times \citep{6145622}. Genetic algorithms, with their population-based approach and genetic operations, may require more computational complexity to evaluate and evolve multiple solutions in parallel \citep{8865431}. However, Tabu search's local search approach and memory-based mechanisms can efficiently explore the neighborhood and avoid revisiting local optima, making it suitable for handling large search spaces \citep{8932381}. Motivated by these, a TS-based LED selection scheme is proposed to maximize the sum secrecy rate of VLC systems, which has not been investigated yet.

 In this paper, we investigate a multi-UEs VLC channel that comprises multiple LEDs with the objective of transmitting confidential messages to UEs in the presence of an active Eve. Assuming that UEs and active Eve send feedback through uplink media, the CSI of UEs and Eve are known by the center control unit. In the center control unit, UE selects an LED based on an algorithm or strategy. The primary goal of the study is to improve the secrecy performance by assigning a distinct LED to each UE while taking Eve's presence into account. To achieve this goal, we formulate a sum secrecy rate maximization problem with integer variables, which is non-convex in nature, and introduce a tabu search (TS)-based algorithm that provides a sub-optimal solution with low complexity. Furthermore, we design three simple LED selection strategies to reduce the complexity further. Simulation results and comparisons of the proposed algorithm with three simple strategies and global search are provided to demonstrate the effectiveness of the proposed algorithm.

  The rest of the paper is organized as follows: In Section II, we describe the system model for the downlink transmission of the multi-UEs VLC system. Section III presents the TS-based LED selection algorithm and three strategies while providing the complexity and convergence analysis. In Section IV, we offer numerical simulation results and comparisons. Finally, we conclude the paper in Section V.

\begin{figure}
	\centering 
	\includegraphics[width=0.5\textwidth]{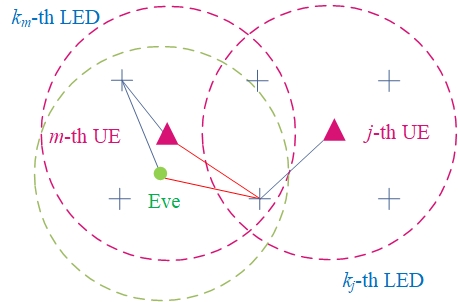}	
	\caption{System model} 
	\label{fig: systemmodel}%
\end{figure}

\section{System model}
There are $M$ UEs and $K$ LEDs in an interior setting with the presence of an Eve. We assume that the connections between LEDs and UEs form one-to-one links. $\mathbf{x}= \left[ {k_1, k_2, k_m, \cdots ,k_M} \right]$ is an index vector that indicates the connection between LEDs and UEs, where ${k_m} \in {\mathbf A}$ is the index of LED that is connected to $m$-th UE, and $\mathbf A= \{1,~2,~k,\cdots ,~K \}$ denotes a set that contains the index of every LED. Once the index vector $\mathbf{x}$ is determined, each UE receives its confidential signal from a specific LED. LEDs that are not selected by any UE are only used for illumination, not for transmitting signals.
However, due to the broadcasting nature of VLC, UE also receives signals from nearby LEDs which are selected by other UEs. For example, as shown in Fig.1, $m$-th UE selects $k_m$-th LED, and $j$-th UE selects $k_j$-th LED.  The $m$-th UE receives two different signals in its reception range. The signal from the $k_m$-th LED is confidential information, whereas the signal from the $k_j$-th LED is considered user interference. When Eve intends to eavesdrop on the $m$-th UE, the signal from the $k_j$-th LED is also regarded as interference. In the meantime, $j$-th UE only receives its own signal from the $k_j$-th LED since no LED selected by other UEs is within its reception range.

In the general case, assuming the index vector $\mathbf{x}$ is fixed, the received signal at $m$-th UE is formulated as follows:  
\begin{align} \label{UE_recSignal}
\text{y}_m = {h_{k_m,m}}{s_m} +\sum\limits_{j=1, ~j \ne m}^{{M}}{{h_{k_j,m}}{s_j}} + n_m,
\end{align}
where $\mathbb{E}\{s_{m}^2\}=\mathbb{E}\{s_{j}^2\}=1$ and $\mathbb{E}\{s_{m}\}=\mathbb{E}\{s_{j}\}=0$. The channel gain from $k_m$-th LED to the $m$-th UE is represented as $h_{k_m,m}$; $s_m$ is the information signal for the $m$-th UE; the channel gain from $k_j$-th LED to the $m$-th UE is $h_{k_j,m}$; $s_j$ is the information signal for $j$-th UE and the part of interference signals for the $m$-th UE. If the $k_j$-th LED falls outside the $m$-th UE's reception range, the channel gain is zero, i.e. $h_{k_j,m} = 0$.  The noise source inside the receivers' circuit is mainly dominated by thermal noise and shot noise. These are modeled as additive zero-mean Gaussian noise, i.e., $n_{m}\sim {\mathcal{N}} (0,\xi^2) $.
 The channel gain from the $k_m$-th LED to the $m$-th UE can be expressed as \citep{Obeed2019}
\begin{equation}\label{channel_gain}
h_{k_m,m}=\frac{(\gamma+1) A_{R}}{2\pi d_{k_m,m}^{2}}\cos(\phi_{k_m,m} )^{\gamma}\cos(\varphi_{k_m,m} )f(\varphi_{k_m,m})g_{\text{of}},
\end{equation}
where $\gamma=\frac{-1}{\log_{2}\cos{\phi_{1/2}}}$ denotes the Lambertian emission order, $\phi_{1/2}$ and $A_{R}$ are the half-intensity radiation angle and the area of receiver's photodiode (PD), respectively, $d_{k_m,m}$ is the Euclidean distance from the $k_m$-th LED to $m$-th UE, $\phi_{k_m,m}$ and $\varphi_{k_m,m}$ are the irradiance angle and incidence angle, respectively, $g_{\text{of}}$ is the gain of the optical filter, and $\text{f} (\varphi_{k_m,m})$ denotes the gain of the optical concentrator which is given by \citep{Obeed2019}
\begin{equation}\label{FOV}
\text{f}(\varphi_{k_m,m}) =
\begin{cases}
\frac{q^2}{\sin^2(\varphi_{k_m,m})} & 0\leq \varphi_{k_m,m}  \leq \Theta,\\
0 & \varphi_{k_m,m} \geq \Theta,
\end{cases}       
\end{equation}
with $q$ and $\Theta$ as the refractive index and the field-of-view (FoV) of the PD used at the $m$-th UE's side, respectively. The extent of FoV dictates the reception range for stationary UEs. The SINR of the $m$-th UE is organized as follows
\begin{align} \label{UE_SINR}
    \text{R}_m = \frac{1}{2}\log_2 \left( {1 + \frac{e}{2\pi}\frac{{{{\left| {{h_{k_m,m}}} \right|}^2}}}{{\sum\limits_{j=1, ~j \ne m}^M {{{\left| {{h_{k_j,m}}} \right|}^2}+ {\xi ^2}} }}} \right).
\end{align} 
Assuming that Eve overhears the confidential signal for the $m$-th UE, which is transferred from the $k_m$-th LED, the received signal at Eve is formulated as 
\begin{align} \label{Eve_recSignal}
\text{ye} = {g_{k_m}}{s_m} + \sum\limits_{j=1, ~j \ne m}^{M} {{g_{k_j}}{s_j}}+ n_e,
\end{align} 
where $g_{k_m}$ is the channel gain from the $k_m$-th LED to Eve; $s_m$ is the confidential signal for the $m$-th UE and also useful signal for Eve; $g_{k_j}$ is the channel gain from $k_j$-th LED to the Eve; $s_j$ is the confidential signal for the $j$-th UE and the part of interference signal for Eve. Thus, the SINR of Eve is organized as follows
\begin{align} \label{Eve_SINR}
\text{Re}= \frac{1}{2}\log_2 \left( {1 + \frac{e}{2\pi}\frac{{{{\left| {{g_{k_m}}} \right|}^2}}}{{\sum\limits_{j=1, ~j \ne m}^{M} {{{\left| {{g_{k_j}}} \right|}^2} + {\xi ^2}} }}} \right).
\end{align} 
 Based on the lower bound of the channel capacity considered in \citep{Yapici2019}, the achievable secrecy rate of the $m$-th UE is given by substituting \eqref{UE_SINR} and \eqref{Eve_SINR} to the following equations: 
 \begin{align} \label{SR}
\text{SR}_m  = [\text{R}_m  - \text{Re} ]_ + .
\end{align} 

The sum secrecy rate maximization problem can be modeled by 
\begin{subequations}\label{orignal_problem}
	\begin{align}
	&~ {\mathrm{max}}\sum_{m=1}^{M} \text{SR}_{m} \label{orignal_problem_a}, \\ 
	\text{s.t.}  &~ \mathbf{x}= \left[ {k_1, k_2, k_m, \cdots ,k_M} \right]\in \mathbb{N}^M, k_m\in  \mathbf A,~~\forall m,    
	\end{align}
\end{subequations}
 The above problem is an integer programming problem, and it involves searching for an optimal index vector $\mathbf{x}$. The discrete nature of the solution space and the non-convexity of the problem are two of the main challenges in solving this problem. The former makes it impossible to find the optimal solution through simple interpolation or approximation, while the latter means there are many local optimal solutions, making it challenging to find the global optimal solution. 

\section{Proposed Algorithm and Strategies}
\subsection{Tabu search-based LED selection algorithm}
In this section, we present the proposed TS-based algorithm for LED selection. The proposed algorithm involves a strategy of connecting the optimal LED for each UE. TS is a metaheuristic optimization algorithm that is widely used to solve complex optimization problems. The algorithm was first introduced in \citep{glover1989tabu} and has since been extensively studied and applied to various fields \citep{TabuEconomic, Srinidhi2011}. In \citep{TabuEconomic}, an improved TS algorithm is developed for a nonlinear economic dispatch problem that the classic Lagrange-based algorithms failed to solve. \citep{Srinidhi2011} proposed a layered detection approach in conjunction with TS and showed that it works very well in terms of both performance as well as complexity in MIMO systems with a large number of antennas.

The TS algorithm begins with an initial solution, defines a set of neighboring solutions based on a neighborhood criterion, and moves to the best one among the neighboring solutions. This process remains a certain number of iterations, after which the algorithm is stopped and the best solution of all the iterations is declared as the final result. In defining the neighborhood of the solution in a given iteration, the algorithm attempts to avoid cycling by prohibiting the moves to the solutions of the past few iterations. The solutions of the past few iterations are recorded in the ‘tabu list’ and parameterized by the ‘tabu period’, which is changed depending on the number of repetitions of the solutions that are marked in the search path. The performance of TS is highly dependent on the selection of its parameters, such as definitions of neighborhood, tabu list, and tabu period.

 \textit{Neighborhood Definition}:  Let ${{\bf{x}}^{(l)}} = \left[ {k_1^{(l)},k_2^{(l)}, \cdots,k_M^{(l)}} \right]$ denotes an index vector belonging to the solution space in the $l$-th iteration, where ${k_m} \in {\mathbf A}$  is the index of LED that is selected by the  $m$-th UE. The neighborhood of the ${k_m}$-th LED for the $m$-th UE is chosen based on ${\mathbf B}_m$. ${\mathbf B}_m$ is a fixed subset of $\mathbf A$, and it contains the indices of every reachable LED for $m$-th UE. As long as $m$-th UE does not move, ${\mathbf B}_m$ is a fixed subset. The cardinality and members of this set are not the same for different UEs, which depends on their FoV and position. Note that the maximum and minimum values of the cardinality are $K$ and 1, respectively.
We refer to a neighbor vector ${{\bf{z}}^{(l)}}(q) = \left[ {z_1^{(l)}(q),{\rm{ }}z_2^{(l)}(q),{\rm{ }} \cdots ,{\rm{ }}z_M^{(l)}(q)} \right]$  as the $q$-th neighbor of ${{\bf{x}}^{(l)}}$, where its $m$-th element  $z_m^{(l)}(q)$ is the  $q$-th  neighbor of  $k_m^{(l)}$,  $q = 1 \cdots M$.  The $z_m^{(l)}(q)$ -th LED is the  $q$-th neighbor of the $k_m^{(l)}$-th LED for the $m$-th UE, whose formula is written as follows: 
\begin{align}
z_m^{(l)}(q) = \omega (k_m^{(l)}), \forall q = 1 \cdots M. 
\end{align}
where ${\omega}(k_m^{(l)})$ represents that the $z_m^{(l)}(q)$ is randomly selected from ${{\mathbf B}_m}$ but not including $k_m^{(l)}$, i.e., $z_m^{(l)}(q)\in {{\mathbf B}_m}\backslash k_m^{(l)}$.  We choose $M$ neighbor vectors that differ from a given vector ${{\bf{x}}^{(l)}}$ in the solution space. These $M$ neighbor vectors are the neighborhood of the given vector ${{\bf{x}}^{(l)}}$.  An operation on ${{\bf{x}}^{(l)}}$ which gives  ${{\bf{x}}^{(l+1)}}$ belonging to the neighborhood of  ${{\bf{x}}^{(l)}}$ is referred to as a move. The algorithm is said to execute a move if ${{\bf{x}}^{(l + 1)}} = {{\bf{z}}^{(l)}}(q)$. We note that the number of candidates to be considered for a move is equal to $M$ in any of the iterations. 

 \textit{Tabu List}: A tabu list $\mathbf{T}$ of size $K \times M$ is the matrix whose entries denote the tabu values of moves. The tabu value of a move means that the move cannot be considered for that many successive iterations. There are $M$ rows in $\mathbf{T}$, where each row represents the indices of UEs from 1 to $M$. The $K$ columns of the $\mathbf{T}$ matrix correspond to the indices of LEDs from 1 to $K$. In other words, the $(k_m, m)$-th entry of the tabu list corresponds to the connection between the $k_m$-th LED and the $m$-th UE. The entries of the tabu list are updated in each iteration, and they are used to determine the direction in which the search proceeds. 
 
 \textit{Tabu Period}: Tabu period, a non-negative integer parameter, is defined as follows: if combinations between LEDs and UEs are accepted as the next move in an iteration, it will remain as tabu for $M$ subsequent iterations until the move becomes a better solution. 
 
 \textit{TS Algorithm}: Let ${\bf{\alpha }}^{(l)}$ be a solution that has the best sum secrecy rate found till the $l$-th iteration. The algorithm starts with an initial solution vector ${{\bf{x}}^{(0)}}$. Set ${{\bf{\alpha}}^{(0)}} = {{\bf{x}}^{(0)}}$. Note that different initial solutions or random choices made during the search process can lead to different results, which can be difficult to replicate or compare. All the entries of the tabu list are set to zero. A binary flag is used to indicate whether the algorithm has reached a local maximum or not. The following steps 1) to 3) are performed in each iteration. Consider $l$-th iteration in the algorithm, $l\geq 0$. 
 
  Step 1): The sum secrecy rate of the neighbor vector ${\text {SR}}({\bf{z}}^{(l)}(q))$  are computed. Let 
  \begin{align} \label{neighbor}
 {\bf{z}}^{(l)}(q^*) = \arg \mathop {\max }\limits_q {\mathop{\text {SR}}\nolimits} 
 ({\bf{z}}^{(l)}(q)), \forall q= 1 \cdots M.
\end{align}     
The neighbor vector  ${\bf{z}}^{(l)}(q^*)$  is accepted as the next move if any one of the following two conditions is satisfied:
 \begin{subequations} \label{conditions}
 \begin{align}
&~{\mathop{\text {SR}}\nolimits} ({\bf{z}}^{(l)}({q^*})) > {\mathop{\text {SR}}\nolimits} ({\bf{\alpha }}_{}^{(l)}),\\
&~{\bf{T}}({z_m}^{(l)}({q^*}),m) = 0, \forall m= 1 \cdots M. 
 \end{align}
 \end{subequations}
The first equation indicates that the secrecy performance of the neighbor vector ${\bf{z}}^{(l)}(q^*)$ surpasses that of the best index vector ${\bf{\alpha }}^{(l)}$ up till the $l$-th iteration. The second formula denotes that the $({z_m}^{(l)}({q^*}), m)$-th entry in the tabu list is marked as zero, which means that the connection between the ${z_m}^{(l)}({q^*})$-th LED  and the $m$-th UE is not forbidden. If this neighbor is not accepted as the next move (i.e. either one of the conditions above is not satisfied), find  the neighbor vector ${\bf{z}}^{(l)}(q^{**})$  with the second best sum secrecy rate, such that 
\begin{align}
{\bf{z}}^{(l)}(q^{**}) = \arg \mathop {\max }\limits_{q: {\rm{ }}{q^{**}} \ne {q^{*}}} {\mathop{\text {SR}}\nolimits} ({\bf{z}}_{}^{(l)}(q)).   
\end{align}
Next, check for acceptance of  ${\bf{z}}^{(l)}(q^{**})$  by substituting to \eqref{conditions}. If this one also cannot be accepted, repeat the procedure for the neighbor vector ${\bf{z}}^{(l)}(q^{***})$ with the third-best sum secrecy rate, and so on. If an acceptable move is available, proceed to Step 2. Otherwise, advance to Step 3. Continue this process until an acceptable move is made.

Step 2): Let  ${\bf{z}}^{(l)}(\hat q)$ be a neighbor vector for which the move is permitted. Make
\begin{align}
  &{{\bf{x}}^{(l + 1)}} = {\bf{z}}^{(l)}(\hat q),\\
  &{\bf{\alpha }}^{(l + 1)} = {\bf{z}}^{(l)}(\hat q), 
\end{align}
where $m$-th element ${z_m}^{(l)}({\hat q})$ of the neighbor vector represents the index of the $\hat q$-th neighbor LED for the $k_m$-th LED, where $m$ ranges from 1 to $M$. It should be noted that the best permissible neighbor vector is selected as the solution vector for the subsequent iteration. Next, this neighbor vector is also assigned to the best solution ${\bf{\alpha }}^{(l + 1)}$ in $(l+1)$-th iteration.

  Step 3):  All the non-zero entries of the tabu list are minus by one, which is updated as follows: 
\begin{align}
  &{\bf{T}}(k_m,m) = \max \{ {\bf{T}}(k_m,m) - 1,0\} ,~\forall m= 1 \cdots M, {\rm{ }}k_m = 1 \cdots K,  \\
  &{\bf{T}}({z_m}^{(l)}({\hat q}),m) = M,~\forall m= 1 \cdots M.
\end{align}
Furthermore, the connection between the ${z_m}^{(l)}({\hat q})$-th LED and the $m$-th UE should be prohibited for several upcoming iterations to prevent cycling. Thus, the $({z_m}^{(l)}({\hat q}),m)$-entry of the tabu list is assigned by the tabu period, denoted as $M$. The algorithm terminates in Step 3 if the following stopping criterion is satisfied, else goes back to Step 1.

   \textit{Stopping Criterion}: The algorithm described above is stopped if the maximum number of iterations is reached. Also, if the total number of repetitions of the current solution is greater than a threshold and this solution is a local maximum, the algorithm is stopped. The average complexity of the algorithm is $O(MK)$. 
  
\subsection{Simple LED selection strategies}
  Three simple LED selection strategies are proposed as benchmarks in this scenario, which require no iteration and have lower complexity. 

\textit{Random strategy}: In this strategy, each UE is assigned to a random LED  within its reception range. The solution vector, denoted as $\mathbf{x}= \left[ {k_1, k_2, k_m, \cdots ,k_M} \right]$ , represents the assignment of LEDs to UEs, where $k_m$ is the index of the LED that $m$-th UE randomly chooses from the set ${\mathbf B}_m$. Note that the random strategy doesn't consider any optimization criteria or system constraints. If any LED  is already connected to a UE, then it cannot be selected as the transmitter for other UEs. 

\textit{Channel gain-based strategy}: In this strategy, each UE is assigned to the LED with the highest channel gain, which is similar to the LED selection method proposed in \citep{Cho2018}. The solution vector is denoted as $\mathbf{x}= \left[ {k_1, k_2, k_m, \cdots ,k_M} \right]$ , where $k_m \in {\mathbf B}_m$ is the index of the LED  that provides the highest channel gain to the $m$-th UE.

\textit{Eve's channel gain-based strategy}: In this strategy, each UE is allocated to the LED with the highest channel gain that is greater than Eve's channel gain. If a UE's channel gain is less than Eve's, the corresponding LED is banned from being selected by that UE. Thus, the indices of banned LEDs are eliminated from the set ${\rm B}_m$. Assuming that $\beta$ is the set containing the index of banned LEDs, the solution vector is denoted as $\mathbf{x}= \left[ {k_1, k_2, k_m, \cdots ,k_M} \right]$ , where $k_m \in {\mathbf B}_m \backslash \beta$ is the index of the LED that provides the highest channel gain to the $m$-th UE.

Since the main procedure of the above strategies involves sorting channel gains from LEDs to each UE, the average complexity of the above three strategies is $O(K)$.  

\subsection{Global search} 
The result of the global search is used to measure the optimality of the proposed algorithm and strategies. To achieve this, a list that contains all possible solutions is created, and the best solution is obtained by substituting every solution vector $\mathbf{x}= \left[ {k_1, k_2, k_m, \cdots ,k_M} \right]$  into \eqref{conditions}. Thus, the average complexity of the algorithm is $O({K^M})$.

\subsection{Convergence analysis }

\begin{figure}[h!]
\begin{center}
\includegraphics[width=10cm]{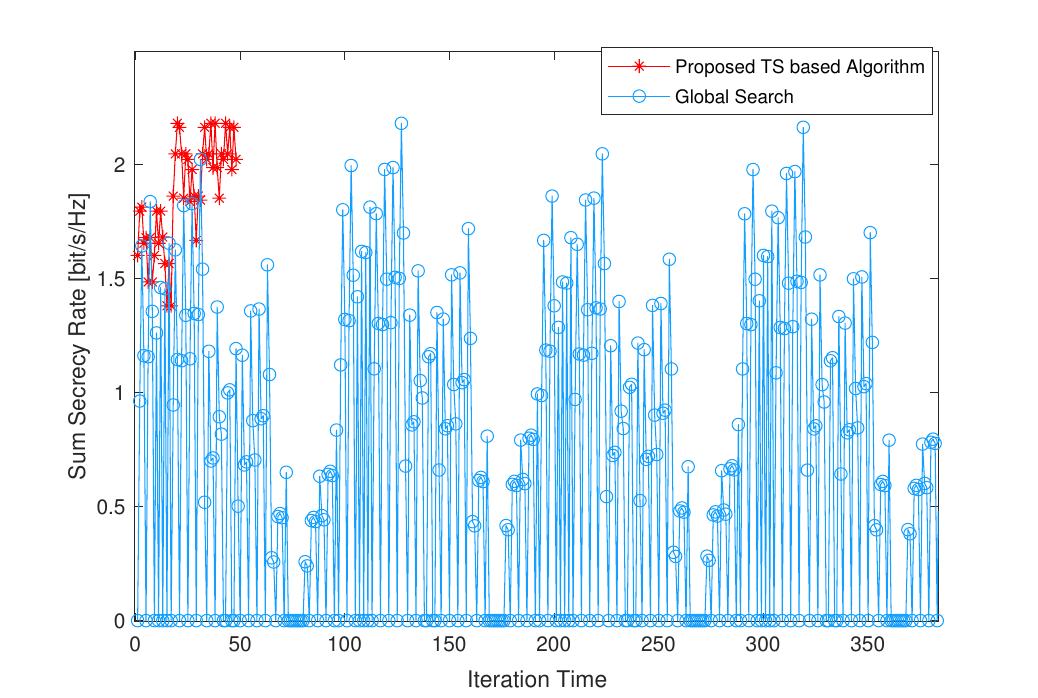}
\end{center}
\caption{Comparison of the proposed algorithm and global search on convergence performance}\label{fig: Convergence}
\end{figure}

Figure \ref{fig: Convergence} illustrates the iteration times of the proposed algorithm in comparison to global search. The proposed algorithm and global search achieve the same optimal value of 2.18 bit/s. However, the proposed algorithm converges to the optimal value after only 50 iterations, whereas the global search process terminates after examining every possible solution, requiring a total of 350 iterations in this example. 

\section{Numerical Results}

\begin{figure}[h!]
\begin{center}
\includegraphics[width=9cm]{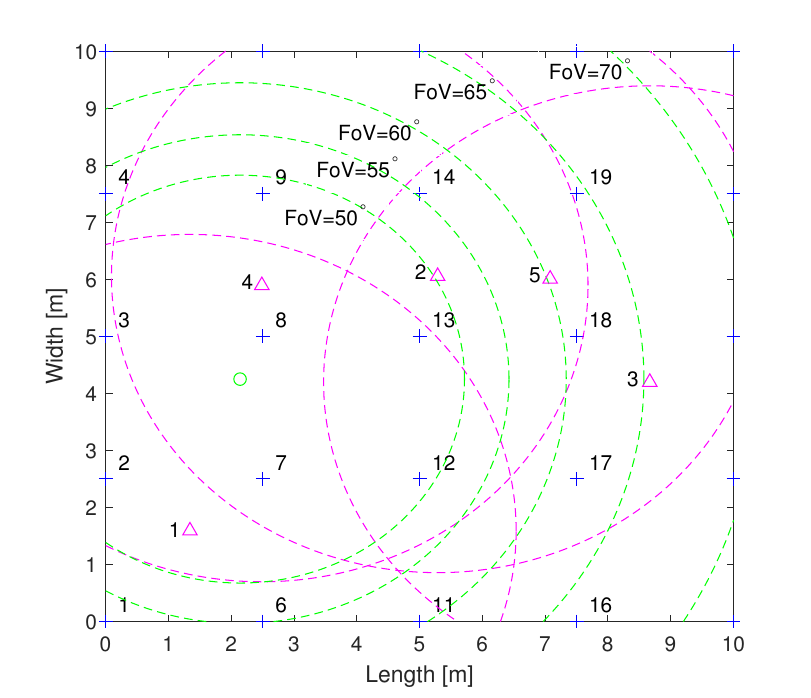}
\end{center}
\caption{Placement of LEDs (blue) and example of a layout for UEs (magenta) and Eve (green)}\label{fig: placement}
\end{figure}

In this section, we study the secrecy performance of the proposed VLC system in a $10\text{~m}\times 10\text{~m}\times 3\text{~m}$ indoor setting. Figure \ref{fig: placement} demonstrates the geometrical configuration of 25 LEDs within an indoor room. Note that the simulations below employ a Monte Carlo-based approach, and the secrecy performance is determined by averaging the results from 500 different instances. In each instance, the coordinates of the UEs and Eve are randomly generated using a uniform distribution within the interval of $[0,10]$. The parameters used in the following simulations are shown in Table 1.

\begin{table}[ht]
	\caption{Simulation Parameters.}
	\centering 
	\begin{tabular}{c| c} 
		\hline 
		Parameter & Value\\ [0.5ex]
		\hline  
		Average electrical ambient noise ($\xi$) & -98 dBm \\ 
		Lambertian emission order $(\beta)$ & 1\\
		Half-intensity  radiation angle $(\theta_{1/2})$  & $60^{\circ}$ \\ 
		PD surface area $(A_R)$ & 1 $\text{cm}^{2}$ \\ 
		Optical filter gain  $(g_{of})$ & 1 \\ 
		Maximum power of a LED & 23 dBm \\  
		PD FoV $(\Theta)$ & $50^{\circ}$ \\
		Refractive index ($q$) & 1.5 \\
		[1ex] 
		\hline 
	\end{tabular}
	\label{parameters} 
\end{table}

Figure \ref{fig: placement} shows a layout example of 5 UEs and one Eve.  In this case, the coordinates of the UEs are (1.34, 1.59, 0.8), (5.28, 6.05, 0.8), (8.66, 4.19, 0.8), (2.48, 5.89, 0.8), and (7.07, 6.00, 0.8), while the coordinate of Eve is (2.13, 4.25, 0.8). The coordinates of the LEDs are denoted by blue forks. The positions of the UEs are marked with magenta triangles, and their respective coverage areas are denoted by magenta circles. The green circle represents Eve's location, and its coverage areas are denoted by green dashed circles. 

\begin{figure}[h!]
\begin{center}
\includegraphics[width=9cm]{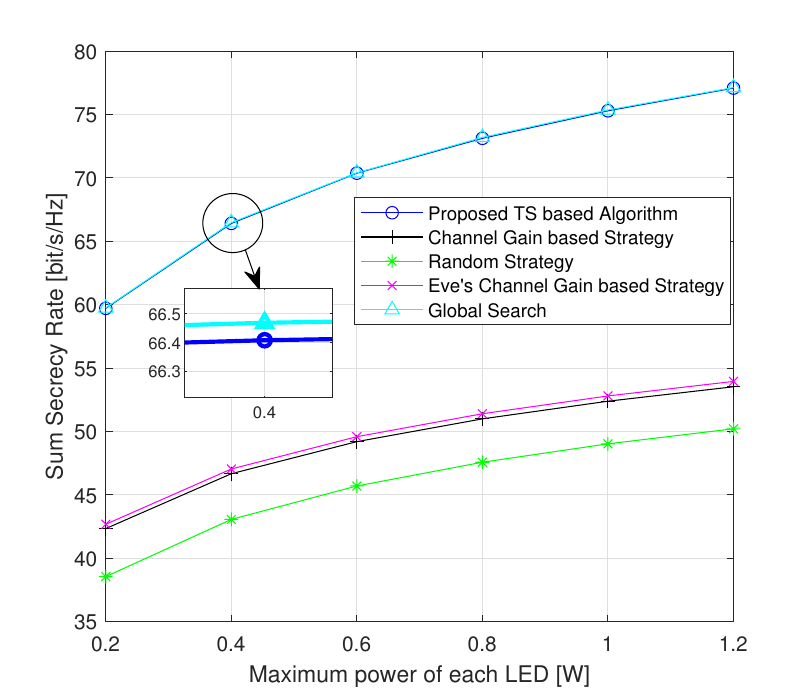}
\end{center}
\caption{Average sum secrecy rate with respect to the maximum power of each LED (randomly generated positions of UEs and Eves for each instance)}\label{fig: SR_P}
\end{figure}

Figure \ref{fig: SR_P} displays the changes in the sum secrecy rate as a function of the maximum power of each LED. The This improvement in secrecy rate is attributed to the increased signal strength transmitted by the LED. When the maximum power of the LED increases, $m$-th UE can receive a stronger confidential signal from its intended LED. If there is no interference signal from nearby LEDs, the increased power directly contributes to an enhancement in the achievable data rate at $m$-th UE. Furthermore,
the results of the global search represent optimal performance in this scenario, and the proposed algorithm's performance is very close to it. The proposed algorithm outperforms three simple strategies, and the performance gap between them becomes more significant with an increasing number of UEs.  Notably, the performance of Eve's channel gain-based strategy is superior to that of the channel gain-based strategy. This is because the latter does not take into account the position of Eve, whereas the former allows UEs to reject the connection of LEDs that provide better channel quality to Eve.

\begin{figure}[h!]
\begin{center}
\includegraphics[width=8cm]{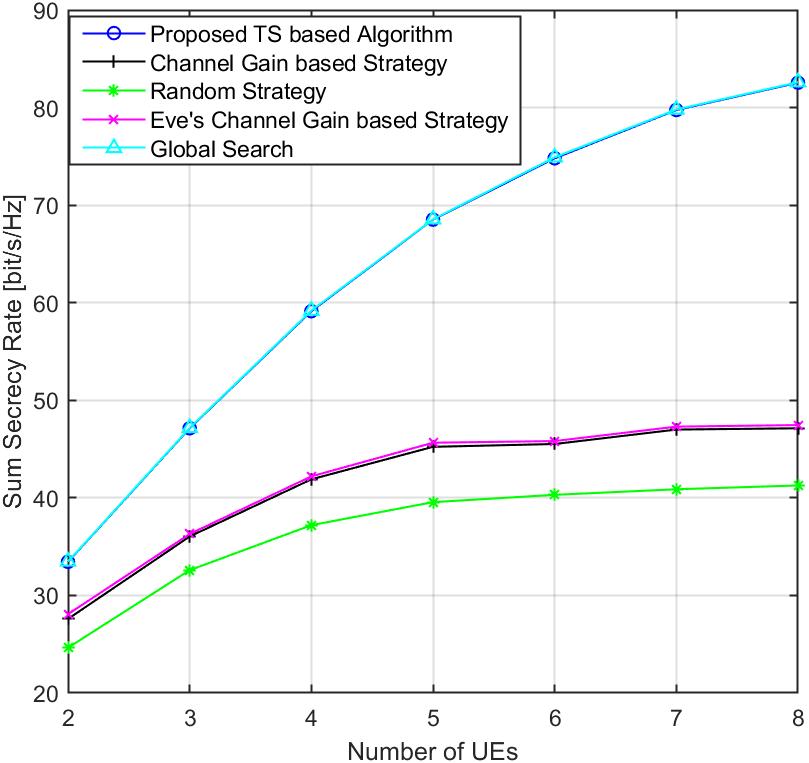}
\end{center}
\caption{Average sum secrecy rate with respect to the total number of randomly placed UEs}\label{fig: SR_NUmUE}
\end{figure}

Figure \ref{fig: SR_NUmUE} illustrates the variation in sum secrecy rate with respect to the number of UEs. As the number of UEs increases, the sum secrecy rate initially experiences rapid growth. This is due to the increased potential for multi-user diversity. UEs that are in favorable positions, for instance, those far away from Eve, contribute to a higher sum secrecy rate. However, the growth rate slows down as the number of UEs continues to increase. This is attributed to the increased inter-user interference and the limitation imposed by the number of LEDs that can be selected for a larger number of UEs.

\begin{figure}[h!]
\begin{center}
\includegraphics[width=9cm]{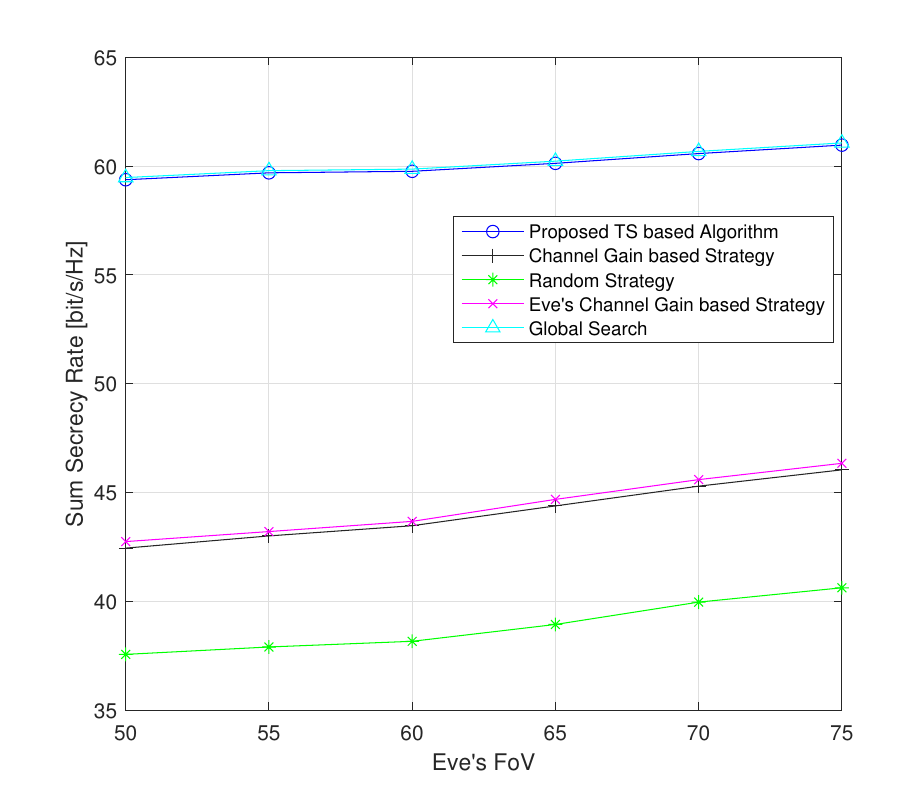}
\end{center}
\caption{Average sum secrecy rate with respect to Eve's FoV (randomly generated positions of UEs and Eves for each instance)}\label{fig: SR_EveFOV}
\end{figure}

Figure \ref{fig: SR_EveFOV} illustrates the performance of the sum secrecy rate as a function of Eve's FoV. The FoV of UEs is 50 degrees. The secrecy performance exhibits a slightly increasing trend with increasing FoV. This is because, at Eve's side, the received power of interference signals grows at a faster rate than that of useful signals. As depicted in Figure \ref{fig: placement}, the green circles represent the detectable areas of Eve. As the FoV increases, the green circles become larger, encompassing more UEs and LEDs, which allows Eve to receive useful signals from more UEs. For example, at a FoV of 70 degrees, Eve begins to receive signals of the 3rd UE. Therefore, compared with the case where its FoV is smaller than 70 degrees, Eve's achievable rate becomes greater than zero when overhearing the 3rd UE. However, if Eve overhears other UEs, their achievable rates become lower due to interference from the 3rd UE.

\begin{figure}[h!]
\begin{center}
\includegraphics[width=9cm]{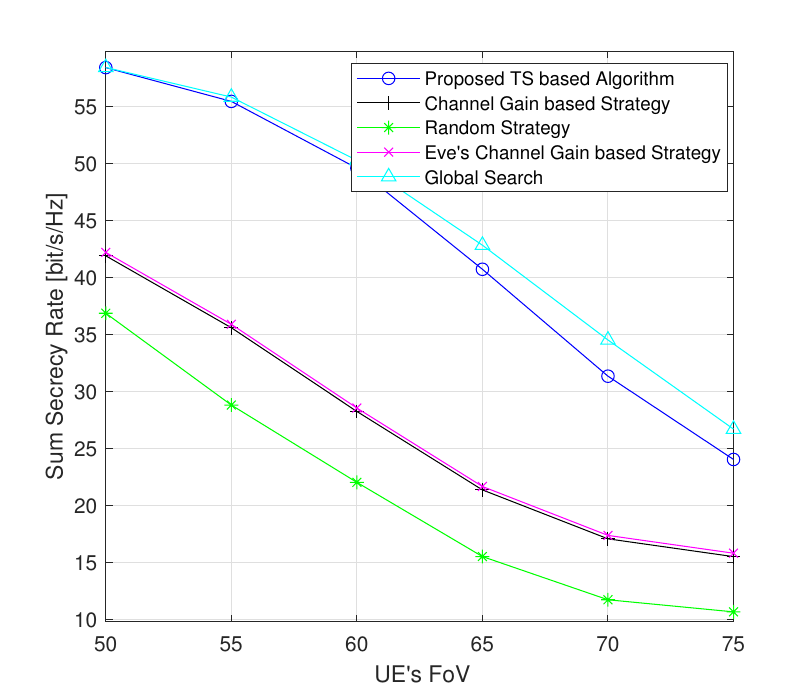}
\end{center}
\caption{Average sum secrecy rate with respect to UEs' FoV (randomly generated positions of UEs and Eves for each instance)}\label{fig: SR_UEFOV}
\end{figure}

Figure \ref{fig: SR_UEFOV} displays the changes in sum secrecy rate as a function of the FoV of UEs, while Eve has a fixed FoV of 50 degrees. As the FoV of UEs increases, the downward trend of the secrecy performance becomes less pronounced. This is because, with a wider FoV, UEs detect signals from more LEDs and receive a greater strength of interference signals. Once the FoV becomes large enough to cover all UEs, the interference remains constant, hence the secrecy performance becomes stable.

\begin{figure}[h!]
\begin{center}
\includegraphics[width=8cm]{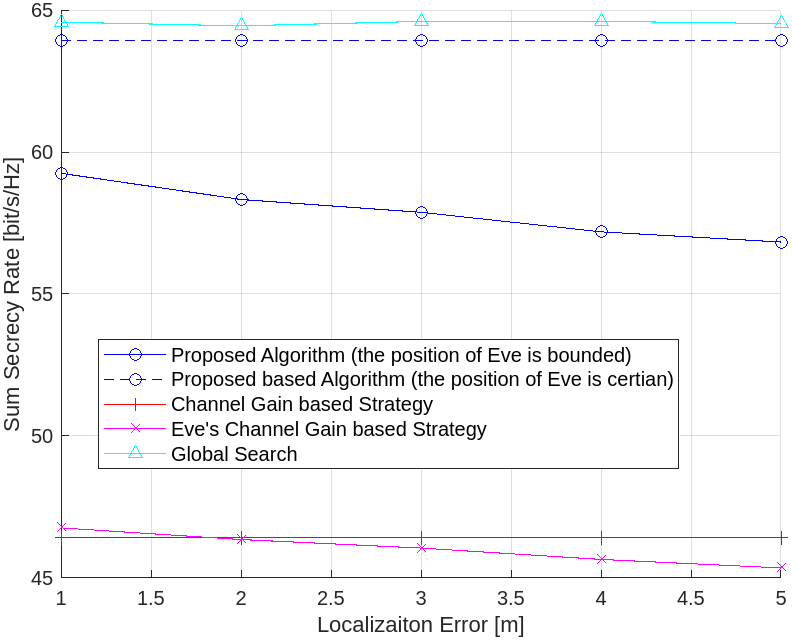} 
\end{center}
\caption{Average sum secrecy rate with respect to the localization error of the Eve (randomly generated positions of UEs and Eves for each instance)}\label{fig: SR_UEEveError}
\end{figure}

Figure \ref{fig: SR_UEEveError} shows the accuracy of the Eve's position influences the performance of the proposed algorithm. Both the secrecy performance of the proposed algorithm and the proposed  Eve's channel gain-based strategy gradually deteriorate as the localization error of the Eve increases. However, the performance of the channel gain-based strategy remains stable, as this strategy assigns the LEDs to the UEs without considering the position of the Eve. 

\section{Conclusions}

In this paper, we present a multi-LEDs VLC system incorporating multiple UEs and a single Eve. Depending on the perfect CSI of UEs and Eve, we propose an LED selection algorithm based on TS to enhance the secrecy performance of the system.  The simulation results demonstrate that the proposed algorithm offers a near-optimal solution with a performance gap of less than 1\%, while achieving faster convergence in comparison to the global search. Additionally, we design three simple LED selection strategies to reduce computational complexity, which both exhibit a performance gap of around 28\% to the global search. Therefore, the proposed algorithm and strategies can be used to balance the trade-off between performance and computational complexity depending on the application requirements. Furthermore, the simulation results show the impact of various system parameters on the secrecy performance. The results indicate that increasing the number of UEs and the maximum power of each LED or reducing UEs’ FoV improves the secrecy performance. On the other hand, augmenting the FoV of Eve has a negligible impact on the secrecy performance. In our future work, we will design a robust TS-based algorithm to enhance security in scenarios where the position estimation of Eve is inaccurate.

\bibliographystyle{elsarticle-harv} 
\bibliography{main}






\end{document}